\documentclass[print]{aa}
\usepackage{graphicx}
\usepackage[varg]{txfonts}
\usepackage{cases}
\usepackage{natbib}
\begin{document}
      \title{Prominence fine structures in weakly twisted and highly twisted magnetic flux ropes}

      \author{J. H. Guo\inst{1,2}, Y. W. Ni\inst{1,2}, Y. H. Zhou\inst{3}, Y. Guo\inst{1,2}, B. Schmieder\inst{3,4}, P. F. Chen\inst{1,2}}

      \institute{School of Astronomy and Space Science, Nanjing University, Nanjing 210023, China \\
      	\email{chenpf@nju.edu.cn; guoyang@nju.edu.cn}
 		\and
 	    Key Laboratory of Modern Astronomy and Astrophysics (Nanjing University), Ministry of Education, Nanjing 210023, China
        \and
        Centre for Mathematical Plasma Astrophysics, Department of Mathematics, KU Leuven, Celestijnenlaan 200B, B-3001 Leuven, Belgium
 	    \and
 	    LESIA, Observatoire de Paris, CNRS, UPMC, Universit\'{e} Paris Diderot, 5 place Jules Janssen, 92190 Meudon, France
        }
    \titlerunning{Prominence fine structures in flux ropes with different twists}
 \authorrunning{Guo et al.}
\date{Received; accepted}

\abstract
{Many prominences are supported by magnetic flux ropes. One important question is how we can determine whether the flux rope is weakly-twisted or strongly-twisted.}
{In this paper, we attempted to check whether prominences supported by weakly-twisted and strongly-twisted flux ropes can manifest different features so that we might distinguish the two types of magnetic structures by their appearance.}
{We performed pseudo three-dimensional simulations of two magnetic flux ropes with different twists.}
{We found that the resulting two prominences differ in many aspects. The prominence supported by a weakly-twisted flux rope is composed mainly of transient threads ($\sim$82.8\%), forming high-speed flows inside the prominence. Its horns are evident. Conversely, the one supported by a highly-twisted flux rope consists mainly of stable quasi-stationary threads ($\sim$60.6\%), including longer independently trapped threads and shorter magnetically connected threads. It is also revealed that the prominence spine deviates from the flux rope axis in the vertical direction and from the photospheric polarity inversion line projected on the solar surface, especially for the weakly-twisted magnetic flux rope.}
{The two types of prominences differ significantly in appearance. It is also suggested that a piling-up of short threads in highly-twisted flux ropes might account for the vertical-like threads in some prominences.}

\maketitle

\section{Introduction}\label{introduction}
Solar prominences, also known as solar filaments, are one of the most fascinating phenomena in astrophysics, which sustain a low temperature ($\sim$7000 K) plasma in the hot solar corona with a temperature of 1--2 million Kelvin. They are about 100 times denser ($10^{10}$--$10^{11}\ \rm cm^{-3}$) than their surroundings, and are believed to be supported by magnetic Lorentz force against gravity. Prominences are usually formed above polarity inversion lines (PILs), which separate positive and negative magnetic polarities. Their threads are slightly skewed from the PIL, implying that prominences are hosted by strongly sheared and/or twisted magnetic structures. Correspondingly, prominences were modeled by either magnetic sheared arcades \citep{Kipp1957} or magnetic flux ropes \citep[MFR;][]{Kuperus1974}. It is noted that the two magnetic configurations in two dimensions (2D) are distinct in the sense that flux ropes have the inverse polarity whereas sheared arcades have normal polarity. In three-dimensional (3D) scenarios, some sheared arcades might also have weak inverse polarity \citep{Devore2000}. In this regard, it was suggested that a magnetic configuration is classified as sheared arcades when the twisted flux is much smaller than the sheared flux \citep{Patsourakos2020}.

The most straightforward method to diagnose the magnetic structure is to measure the magnetic field of prominences based on the Zeeman effect or Hanle effect. The measurements conducted by \citet{Leroy1984} and \citet{Bommier1998} indicated that the majority of prominences belong to the inverse-polarity type, and much fewer belong to the normal-polarity type. Considering that the magnetic measurements of prominences are not routinely available nowadays, an indirect method was proposed by \citet{chen14} to distinguish between sheared arcades and flux ropes based on extreme ultraviolet (EUV) imaging observations only, without the help of magnetic measurements \citep[see][for a schematic]{chen20}. Applying this method to 571 filaments, \citet{Ouyang2017} revealed that $\sim89$\% of the filaments are supported by flux ropes, and $\sim$11\% are supported by magnetic sheared arcades.

However, all the above-mentioned methods cannot tell how strongly the coronal magnetic field is twisted around a prominence, which is crucial in determining whether the prominence might experience kink instability \citep{Hood&Priest1979}. For this purpose, a better and widely-adopted way is to derive the coronal magnetic field surrounding a prominence via nonlinear force-free field (NLFFF) extrapolations based on the vector magnetograms on the solar surface. The NLFFF extrapolations have actually made great contributions to the understanding of the magnetic structures of prominences \citep{wieg21}. For example, \citet{Guo2010} found that an active-region prominence is partly supported by a twisted flux rope and partly by a sheared arcade; \citet{Jiang2014} constructed a twisted flux rope model for a large-scale prominence. Regarding the magnetic twist of extrapolated flux ropes, some prominences were claimed to be supported by weakly twisted flux ropes \citep{Jibben2016, Luna2017}, and others were claimed to be supported by highly twisted flux ropes \citep{Su2015, Guo2019, Mackay2020}. \citet{Guojh20211} found that the mean twist of a flux rope is proportional to its aspect ratio. Among that, the magnetic twists of some large-scale filaments can reach 3 turns \citep{Guo2019, Guojh20211}. Notably, both simulations and observations suggested that the critical twist for the kink instability depends on the flux-rope configuration \citep{Torok2004, Torok2005, Wang2016, Liu2019}. Apart from that, the gravity of the filament material can also suppress the eruption of a flux rope \citep{Fan2018, Fan2020}. Therefore, highly twisted flux ropes may probably exist before the eruption, in particular for large-scale quiescent prominences located in weak-field regions. The diverse results raise a concern whether the magnetic flux ropes are always weakly twisted or can be highly twisted before the eruption. The uncertainty of the extrapolated coronal magnetic configuration results from the weakness of the extrapolation method: First, the NLFFF extrapolation is an ill-posed problem \citep{low90}, and the existence of inevitable magnetic tangential discontinuity can hardly be extrapolated \citep{low15}. Second, there exists the 180$^\circ$ ambiguity in the transverse magnetic field measurements, which is crucial in determining whether the core magnetic field of the prominence is a magnetic sheared arcade or a flux rope based on the extrapolated coronal magnetic field. Furthermore, quiescent prominences are usually located in decayed active regions or quiet regions, where the transverse magnetic field is too weak to be measured precisely so far. As a result, the direct NLFFF extrapolations, such as the optimization method \citep{Wheatland2000, Wiegelmann2004} and the magneto-frictional method \citep{Yang1986, Guo2016a, Guo2016b}, might be erroneous in modeling reasonable magnetic structures around quiescent prominences. It is noted that the uncertainty in coronal magnetic field extrapolation can be alleviated by combining coronal observations, as used in the flux-rope insertion method \citep{Ballegooijen2004} or the regularized Biot-Savart laws \citep[RBSLs;][]{Titov2018}.

On the other hand, as the building block of a prominence, cold dense threads are believed to trace the local magnetic field lines, so the thread characteristics strongly depend on their supporting magnetic configuration. Therefore, the magnetic configuration of a prominence can be reflected through the morphology and fine structures of the prominence. For example, \citet{Karpen2003} found that the prominences supported by sheared arcades are more dynamic than those supported by flux ropes. \citet{Zhou2014} found that the thread length increases with the length of its supporting magnetic dip and decreases with the dip depth. \citet{Guojh20212} found that the magnetically connected threads in double-dipped flux tubes are usually shorter than independently trapped threads in single-dipped flux tubes. Besides, the magnetic measurement of prominences showed that the mean angle between the prominence spine and magnetic field is about $53^{\circ}$ in the normal-polarity prominences, and is about $36^{\circ}$ in the inverse-polarity prominences \citep{Bommier1994}. Some authors also utilized thread flows to estimate the magnetic twist of a prominence \citep{Vrsnak1991, Romano2003}. Therefore, it would be interesting to investigate whether weakly-twisted flux ropes and strongly-twisted flux ropes would be manifested differently in imaging observations.

In this paper, we simulate the formation of two prominences supported by two flux ropes with different twists, and compare their morphologies. This paper is organized as follows. The numerical setup is introduced in Section \ref{sec:met}, the results are presented in Section \ref{sec:res}, which are followed by discussions and summary in Sections \ref{sec:dis} and \ref{sec:sum}.

\section{Numerical setup}\label{sec:met}

In principle, the problem under study is a 3D one, as done by \citet{Xia2014} and \citet{Xia2016}. However, limited by the current computing power, real 3D simulations can not guarantee high spatial resolution, which is important for simulating the fine structures of prominences. For this purpose, we adopt the pseudo-3D approach as used by \citet{Luna2012}, that is, we assume that the magnetic field remains unchanged during the prominence formation process, and the plasma dynamics in different magnetic flux tubes are completely independent. This approximation is valid when both the plasma $\beta$ (the ratio of gas to magnetic pressures) and plasma $\delta$ (the ratio of gravity to magnetic pressure) are small \citep{Zhou2018}. Therefore, a pseudo-3D simulation is equivalent to a collection of many one-dimensional (1D) hydrodynamic simulations, where the geometries of all the magnetic flux tubes are defined by the 3D force-free magnetic field. Such an approach is practical for other reasons: First, for most of the quiescent phenomena, the magnetic structure can be regarded as quasi-static on the time scale of prominence formation \citep{Martin1998}. Second, the thermal conductivity perpendicular to the magnetic field line is about $10^{12}$ times smaller than that parallel to the magnetic field line \citep{Braginskii1965}, hence neighboring flux tubes can be considered as thermally isolated. The pseudo-3D approach can overcome the drawback of low resolution in 3D full magnetohydrodynamics (MHD) simulations, and has the advantage of the 3D visualization at the same time. Therefore, this method has been widely used to study prominence formation \citep{Karpen2003, Luna2012, Guojh20212}, prominence oscillation \citep{Zhang2012, Zhang2020, Ni2022} and long-period intensity pulsations in coronal loops \citep{Froment2017}.

The 1D hydrodynamic equations, as displayed in our earlier works \citep{Xia2011, Zhou2014, Guojh20212}, are numerically solved with the Message Passing Interface Adaptive Mesh Refinement Versatile Advection Code \citep[MPI-AMRVAC \footnote{\url{http://amrvac.org}};][]{Xia2018, Keppens2020}. Note that the field-aligned thermal conduction and optically-thin radiative cooling are both included in the energy equation. The code enables adaptive mesh refinement, and we use six levels of refinement with 960 base-level grids for each 1D simulation, which leads to an effective spatial resolution ranging from 3.5 to 39.1 km. With the appropriate magnetic field environment, cold material of prominences can be formed under the following models: the evaporation-condensation model \citep{Antiochos1999, Karpen2003, Xia2011, Xia2016, Zhou2020}, the injection model \citep{An1988, Wang1999}, and the levitation model \citep{rust94}. Here we resort to the evaporation-condensation model, where the background heating is the same as in our previous works \citep{Xia2011, Zhou2014}.

For the purpose of pseudo-3D simulations, we should first provide an appropriate 3D magnetic field distribution with a flux rope. Such a magnetic configuration can be realized by simulating the evolution of magnetic arcades through magnetic reconnection driven by vortex and converging flows \citep{Xia2014b, Xia2016, Zhou2018, Luna2012}, flux-rope emergence \citep{Fan2001}, or by constructing theoretical models \citep{td99, Titov2014, Titov2018}. The former two methods need higher computational cost, and are difficult to control the twist of a flux rope. Thus, we choose the third method, that is, constructing an analytical force-free Titov-D\'emoulin-modified (TDm) magnetic flux rope model \citep{Titov2014, Titov2018}. In this model, the mean twist of the magnetic field is proportional to the aspect ratio of the flux rope \citep{Guojh20211}.

\begin{figure}
	\includegraphics[width=7cm,clip]{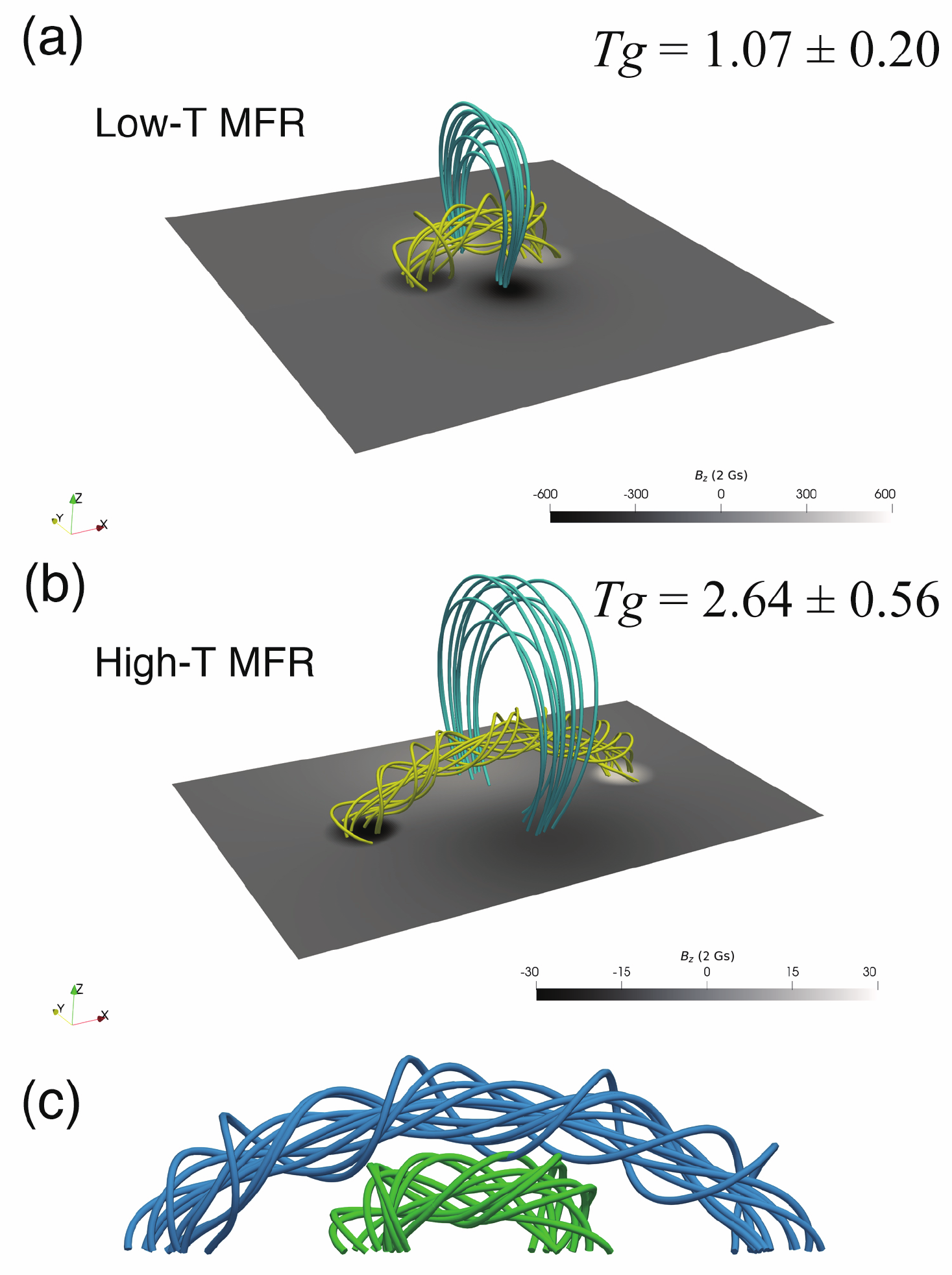}
	\centering
	\caption{Magnetic field models with low and high twists. Panels (a) and (b) represent the low-twist ($|\overline{T_{\rm g}}|=1.07 \pm 0.20$) and high-twist flux rope ($|\overline{T_{\rm g}}|=2.64 \pm 0.56$), respectively. Yellow lines denote the core flux-rope field, and cyan lines denote the background potential field. The twist error is the standard deviation of all sample field lines. Panel (c): Side view along the $y$-axis showing two flux ropes in the same coordinate system, where the green and blue lines represent the low-twist and high-twist flux ropes, respectively. \label{fig1}}
\end{figure}

The details of the TDm model construction are as follows. First, we set two magnetic charges of strength $q$ lying at a depth $d_{q}$ and separated by a distance $2L_{q}$ to construct the background potential field. Second, we set the physical parameters for two flux ropes, that is, the minor radius of the flux rope ($a$), and the major radius of the ring ($R_c$). With that, we can compute the toroidal electric current \citep{Titov2014} and magnetic flux \citep{Titov2018}. Third, we construct the flux rope by the RBSL method with the aforementioned parameters, and embed it into the background field. To quantitatively study the twisting property of the two flux ropes, we calculate the twist number ($T_{\rm g}$) using Formula (12) in \citet{Berger2006}. In the two models, one has a weakly-twisted flux rope, and the other has a highly-twisted flux rope. For the low-twist flux rope model (labeled as Low-T model), $R_{\rm c}=43\ \rm Mm$, $a=20\ \rm Mm$, $d_{q}=16\ \rm Mm$, $q=40\ \rm T \cdot Mm^{2}$ and $|\overline{T_{\rm g}}|=1.07$. The distance between the two footprints of the flux rope, $L_\mathrm{fp}$, is about 80 Mm, and the apex height, $h_\mathrm{ap}$, is about 27 Mm. For the high-twist flux rope model (labeled as High-T model), $R_{\rm c}=170\ \rm Mm$, $a=25\ \rm Mm$, $d_{q}=115 \rm \ Mm$, $q=50\ \rm T \cdot Mm^{2}$ and $|\overline{T_{\rm g}}|=2.64$. Correspondingly, $L_\mathrm{fp}=250\ \rm Mm$, and $h_\mathrm{ap}=55\ \rm Mm$. The mean twists are $1.07\pm 0.20$ and $2.64\pm0.56$ for the two flux ropes, respectively. These quantities are in the typical range of solar prominences \citep{Tanberg1995, Engvold2015}. Figure \ref{fig1} shows the two magnetic field models. It is noticed that the twist of the magnetic configurations in \citet{Luna2012} and \citet{Devore2005} is about 1 turn, which is analogous to our Low-T model.

From each magnetic field model, we select 250 magnetic field lines to perform 1D hydrodynamic simulations, which are roughly uniformly distributed inside the magnetic flux rope. The localized heating is symmetrically imposed at the two footpoints of each flux tube, with the amplitude $E_1= 1.0 \times 10^{-2}$ $\rm erg\,cm^{-3}\,s^{-1}$ \citep{Xia2011}. Similar to \citet{Karpen2003}, the localized heating is ramped up linearly over 860 s and maintained thereafter. Since the field-aligned gravity is not uniform and flux-tube dependent, we compute its distribution according to the field line path, that is, $g_{\parallel}(s)=\ \boldsymbol{g_{\odot}} \cdot \hat{e_{s}}$, where $\boldsymbol{g_{\odot}}=-274\hat{e_{z}} \ \rm m \ s^{-2}$, $s$ is the 1D coordinate along the field line, and $\hat{e_{s}}$ is the unit tangential vector along the field line. Our atmospheric model is the same as in the previous simulations \citep{Zhou2017, Guojh20212}, where the chromospheric temperature is about 6000 K, the density at the bottom is about $1.6 \times 10^{14}\ \rm cm^{-3}$, the coronal temperature is about 1 MK, and the thickness of the chromosphere is about 3 Mm. The atmosphere obtained in this way is not in thermodynamic equilibrium, therefore we relax this initial state to a state both in force and energy equilibria in about 57 min.

\section{Numerical Results}\label{sec:res}
\subsection{Thread formation process and fine structures} \label{sec:NR1}

\begin{figure*}[htbp]
	\includegraphics[width=15cm,clip]{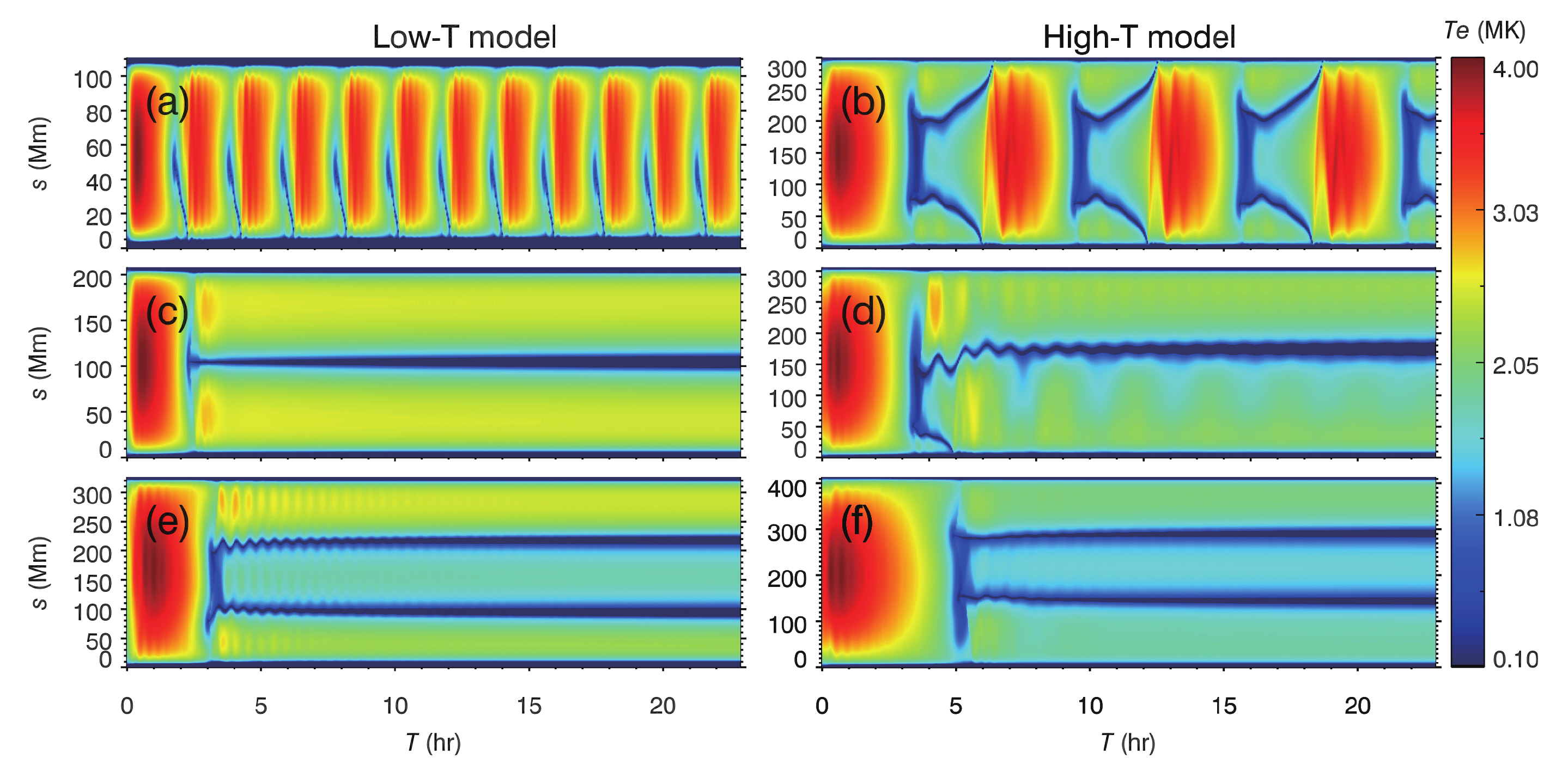}
	\centering
	\caption{Temporal evolution of the temperature distribution along a flux tube in the Low-T mode (left column) and the High-T model (right column), including dynamic threads (panels a and b),  independently trapped threads (panels c and d), and magnetically connected threads (panels e and f). \label{fig2}}
\end{figure*}

Note that although the mean twist is significantly different between the two model flux ropes, both of them have three types of field lines, that is, non-dipped, single-dipped and multi-dipped field lines, with different percentages. Figure \ref{fig2} depicts the temporal evolution of the temperature distributions along three types of flux tubes in each model. The first row, Figures \ref{fig2}a and \ref{fig2}b, shows the dynamic thread formation and evolution processes along the non-dipped flux tubes in Low-T and High-T models. It is seen that the flux tubes without dips experience thermal non-equilibrium cycles, similar to previous works \citep{Karpen2001, Froment2017}, forming dynamic threads moving at a high speed \citep{Karpen2001}. For the dynamic threads in the Low-T model (Fig. \ref{fig2}a), the period of the thread appearance is about 114.7 min, the thread lifetime is about 28.6 min, and the average velocity during the drainage is about 25.2 km s$^{-1}$, which is consistent with observations \citep{Zirker1998, Lin2003, Lin2005}. For that in the High-T model (Fig. \ref{fig2}b), the period of the thermal non-equilibrium cycle is about 378.4 min, the thread lifetime is about 194.9 min, and the average velocity is roughly 7.9 km s$^{-1}$. One can find that the period of the thermal non-equilibrium
cycle increases with the flux-tube length, which is consistent with the results of \citet{Luna2012}. These small dynamic threads, called ``blobs'' in \citet{Luna2012}, are generally formed in non-dipped arcades or shallow-dipped field lines and move downward to the chromosphere, forming counterstreaming flows. The second row of Fig. \ref{fig2} illustrates the evolution of temperature along single-dipped flux tubes in the two models (Figs. \ref{fig2}c and \ref{fig2}d). In the Low-T model, only one thread is formed near the magnetic dip after the localized heating continues for 2.34 hr. The thread is rather stable after quick damping of short-period oscillations, growing with time as the chromospheric evaporation continues. In contrast, in the High-T model, a thread is formed slightly away from the magnetic dip at $t$=3.59 hr, which then oscillates for several periods until it becomes stable at the magnetic dip. Besides, a second thread is formed at $s$= 50 Mm. This thread drains down to the footpoint, forming a dynamic thread as depicted in Fig. \ref{fig2}d. The formation of a second thread is consistent with previous works \citep{Karpen2006, Xia2011, Guojh20212}, if only the field line is long. In particular, longer field lines could delay the onset of catastrophic cooling \citep{Guojh20212}. Therefore, even with the same localized heating, the formation of prominence thread supported by highly-twisted flux ropes needs more time than prominences supported by weakly-twisted flux ropes. The third row of Fig. \ref{fig2} shows the evolution of the temperature along double-dipped flux tubes in the two models, and two threads are formed in each model as seen in panels (e) and (f).

\begin{figure*}[htbp]
    \includegraphics[width=13cm,clip]{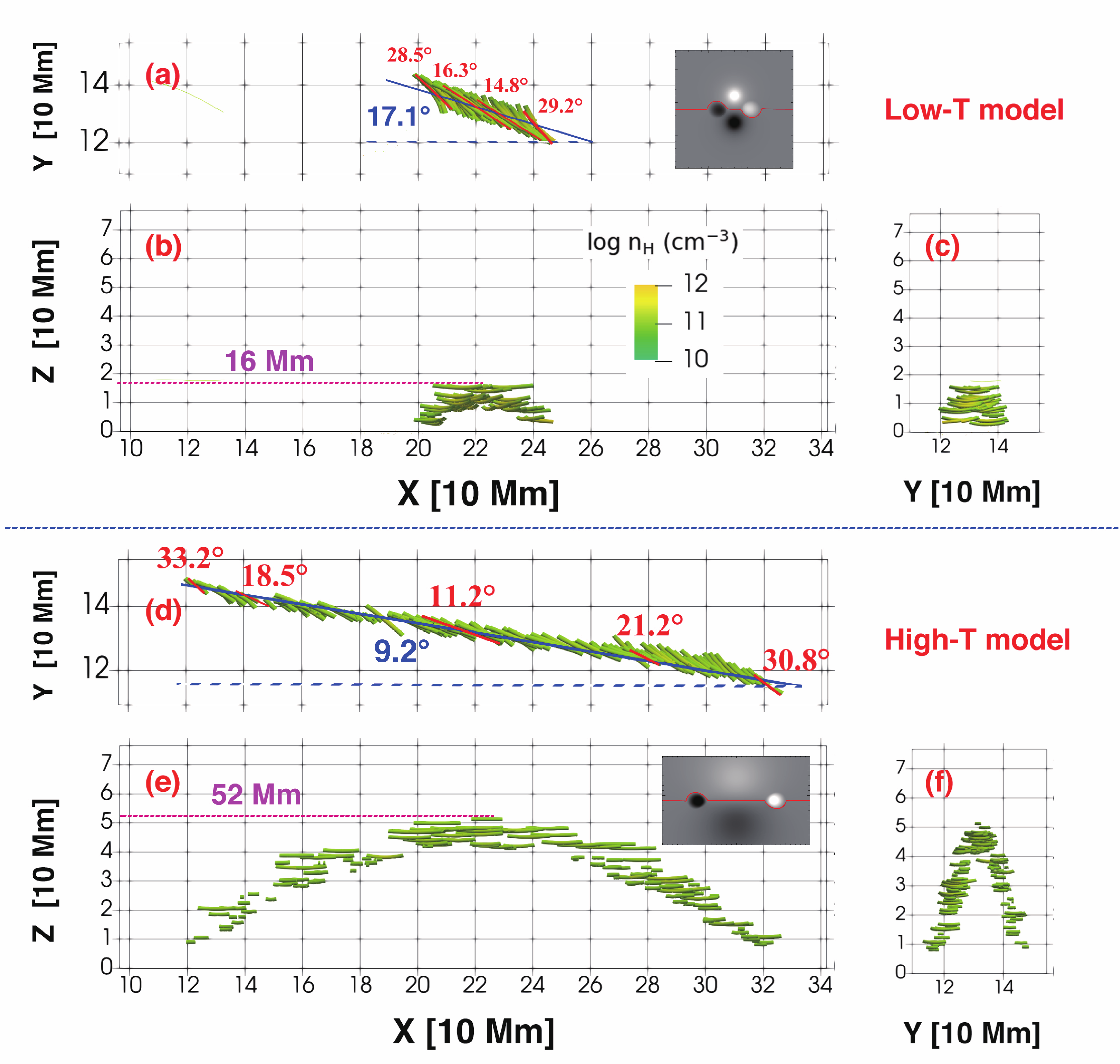}
	\centering
	\caption{Distribution of the prominence threads in different views. Panels (a--c) correspond to the top, side and end views of the Low-T model; Panels (d--f) correspond to the top, side and end views of the High-T model. The blue solid lines in panels (a) and (d) represent the filament spines. The angles marked in blue represent those between the flux rope axes and filament spines, and the angles marked in red represent those between the thread orientations and filament spines. The insets show the maps of magnetic field in the photosphere, where the red solid lines represent the PILs. \label{fig3}}
\end{figure*}

With the simulation results of all the 250 flux tubes in each model at the end of simulations ($t\sim$23 hr), we can compile them and construct a 3D view of each prominence. In this paper, a prominence is called filament when viewed from above, similar to the terminology in observations. Figures \ref{fig3}a--\ref{fig3}c display the distribution of the formed threads in the Low-T model viewed from different perspectives. As viewed from the top, we can see in Fig. \ref{fig3}a that although the photospheric magnetic PIL is along the $x$-axis, the filament spine, as indicated by the solid blue line, is inclined to the photospheric PIL (blue dashed line) by $17.1^\circ$. The filament spine is composed of many threads, and the thread orientation deviates from the filament spine by an angle ranging from $15^\circ$ to $30^\circ$. For the side view along the $y$-axis in panel (b), it is seen that all the threads pile up into an arch-shaped prominence, with the apex at an altitude of 16 Mm. For the end view along the $x$-axis in panel (c), it is seen that the apparent width is quite large, and the two legs are overlapping. For comparison, we show the distribution of the formed threads in the High-T model from different perspectives in Figs. \ref{fig3}d--\ref{fig3}f. For the top view along the $z$-axis in panel (d), we find that the filament spine is more parallel to the photospheric PIL. Moreover, the deviation angle between the threads and the filament spine is minimal near the middle of the flux ropes, and increases toward the footpoints for both of two cases. For the side view, the prominence can approximately outline the path of the flux rope axis in the High-T model (Fig. \ref{fig3}e). Regarding the end view, one can see that the prominence in the High-T model is composed of many short threads compared to the prominence height (Fig. \ref{fig3}f), reproducing a vertical thread-like structure.

\begin{figure*}
	\includegraphics[width=13cm,clip]{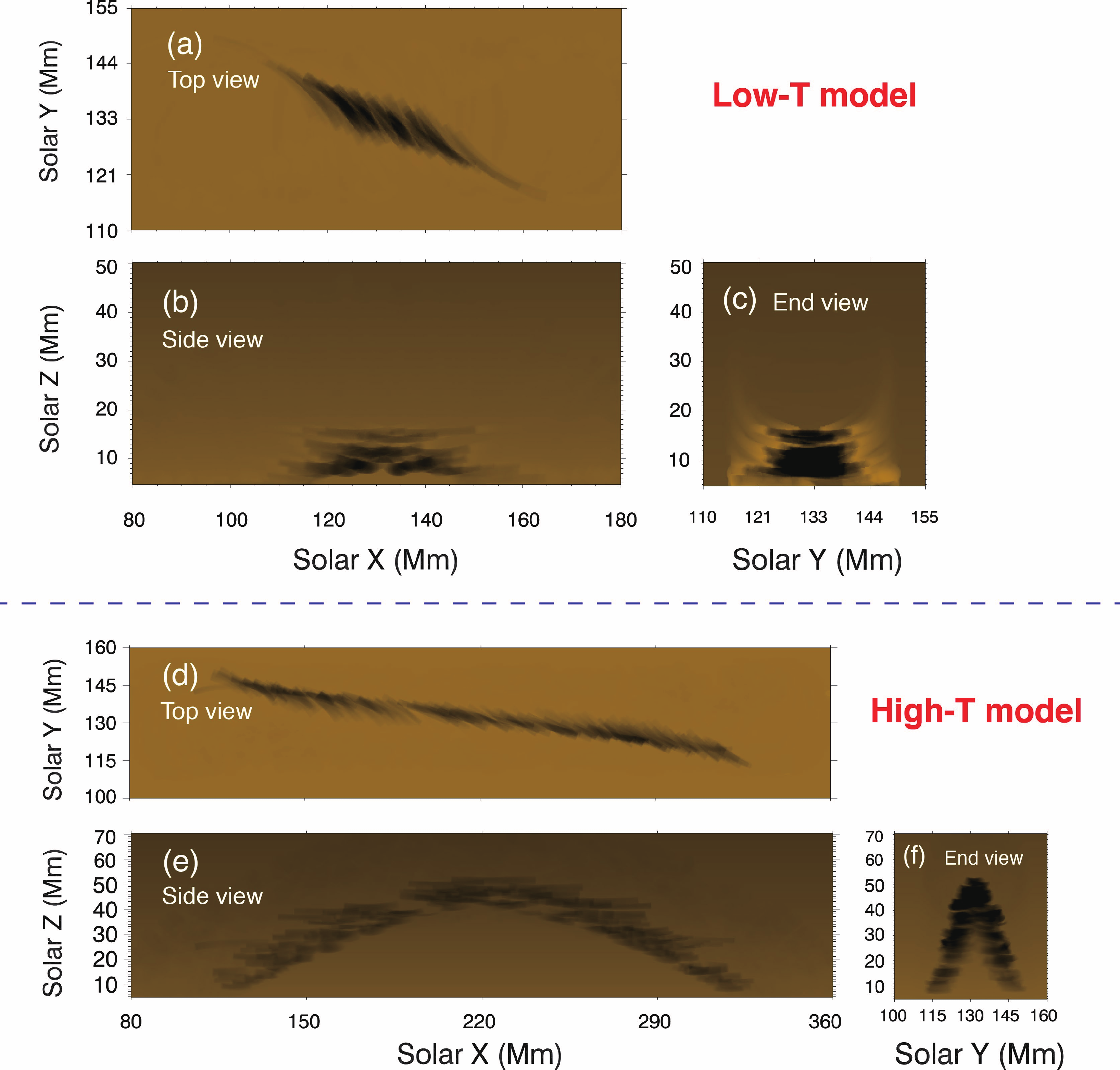}
	\centering
	\caption{Synthesized EUV 171 \AA\ images in the two models. Panels (a--c) correspond to the top, side and end views of the Low-T model; Panels (d--f) correspond to the top, side and end views of the High-T model. \label{fig4}}
\end{figure*}

The cold and dense prominences are optically thick in the EUV range, that is, partial incident light is absorbed through the photoionization of the neutral hydrogen (H \small{I}), neutral helium (He \small{I}), and singly ionized helium (He \small{II}). Therefore, we solve the radiative transfer equation to synthesize the EUV radiation images \citep[see also][]{Zhao2019, Zhou2020, Jenkins2022}, which is different from our previous optically-thin modeling \citep{chen06}. The solution to the radiative transfer equation \citep{Rybicki1986} is as follows:

\begin{eqnarray}
I_{\lambda}(\tau_{\lambda})=I_{\lambda}(0)e^{-\tau_{\lambda}}+\int_{0}^{\tau_{\lambda}}e^{-(\tau_{\lambda}-\tau^{'}_{\lambda})}S_{\lambda}(\tau^{'}_{\lambda})d\tau^{'}_{\lambda} \label{eq1}
\end{eqnarray}
\noindent
where $I_{\lambda}(\tau_{\lambda})$ is the measured specific radiation intensity at the wavelength $\lambda$, $I_{\lambda}(0)$ is the incident intensity in the background, $\tau_{\lambda}$ is the total optical depth, $\tau^{'}_{\lambda}$ is the local optical thickness, $S_{\lambda}=j_{\lambda}/\alpha_{\lambda}$ is the source function, $j_{\lambda}$ is the emission coefficient and $\alpha_{\lambda}$ is the absorption coefficient. Among them, the optical depth and source function are strongly dependent on the relative population of hydrogen and helium, therefore we need to calculate the ionization degree with the Saha equation. Then, the emission coefficient can be calculated according to Formula (24) in \citet{Jenkins2022}, and the absorption coefficient can be obtained from Formula (6) in \citet{Anzer2005}. Finally, similar to \citet{Jenkins2022}, assuming an incident intensity $I_{\lambda}(0)$, the transfer equation can be computed numerically, and the synthesized 171 \AA\ images viewed from different directions are shown in Fig. \ref{fig4}.

For the top views, we find that the overall structure of the filament in the Low-T model looks like a rhombus, whereas that in the High-T model presents stick-like morphology. Regarding the fine structures, it is seen that the filament spine is composed of right-bearing threads, which is in accord with the chirality of the filament and the helicity of the supporting flux rope. Moreover, the edge of the filament spine in the High-T model is serrated while that in the Low-T model is smoother. Comparing the side views of the two models (Figs. \ref{fig4}b and \ref{fig4}e), we find that there are no obvious legs in the prominence supported by a low-twist flux rope. That is to say, these prominences would be manifested as being nearly detached from the solar surface. However, for the prominence supported by a highly twisted flux rope, its legs almost extend down all the way to the solar surface. For the end views along the flux rope axes (Figs. \ref{fig4}c and \ref{fig4}f), cavities all appear in two models on the top of the prominence condensations, which is consistent with 3D full MHD simulations of the flux-rope prominence formation \citep{Xia2014, Xia2016, Fan2018, Fan2019}. Other than that, we also find that the horn-like structures are much more evident in the Low-T model than in the High-T model.

\subsection{Magnetic dip and thread characteristics} \label{sec:NR2}

High-resolution observations revealed that prominences are composed of numerous separate fine-scale threads \citep{Lin2005}, which can be divided into dynamic threads, independently trapped threads and magnetically connected threads \citep{Guojh20212}. These threads were found to move with velocities of roughly 15$\pm$10 km $\rm s^{-1}$ \citep{Engvold2004, Lin2005}. The relationship between threads and magnetic dips is an important issue. Our pseudo-3D simulations provide a chance to study their relationship in detail.

\begin{figure*}
	\includegraphics[width=10cm,clip]{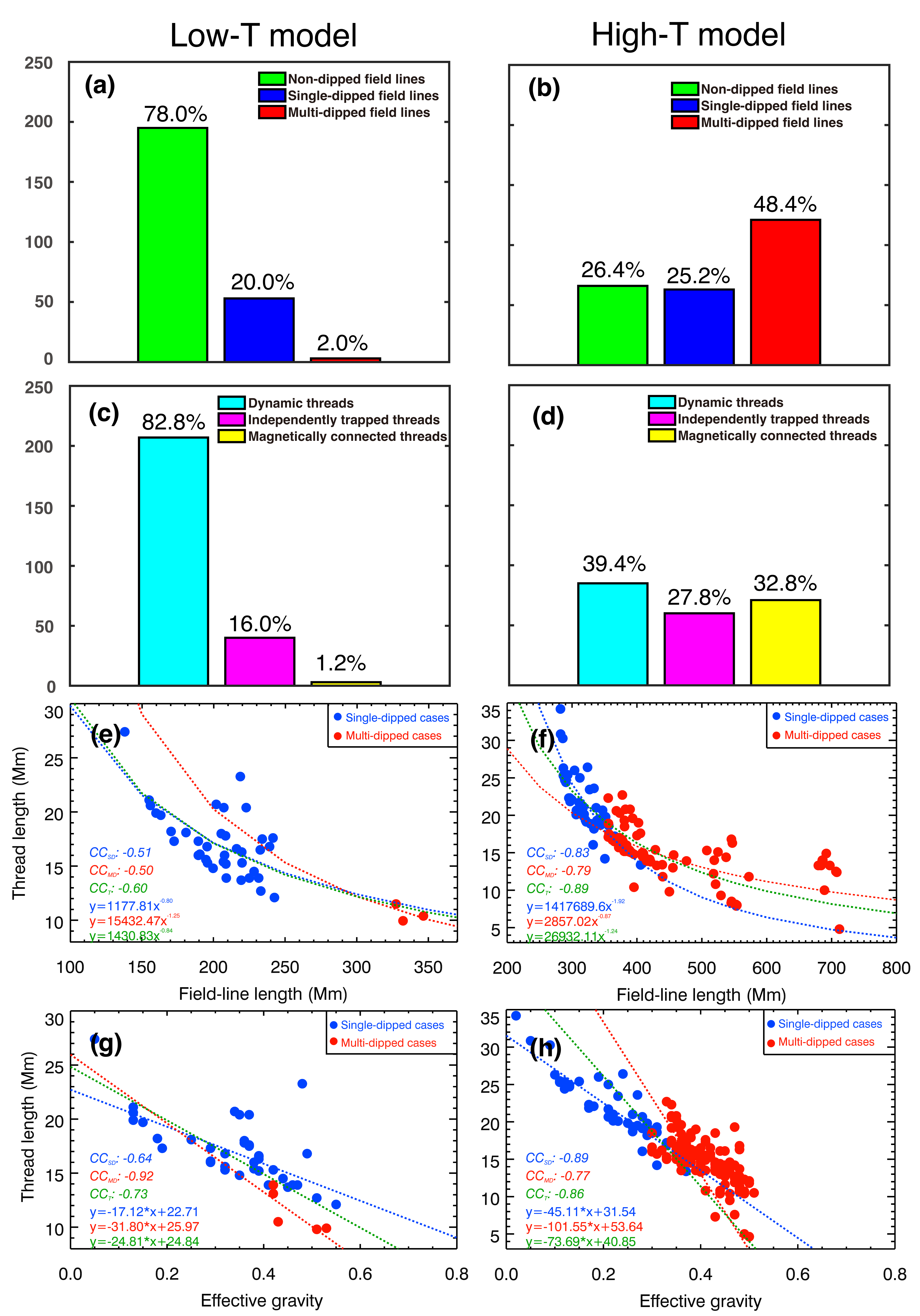}
	\centering
	\caption{Statistical results of magnetic dip and thread characteristics in the Low-T (left column) and High-T (right column) models. Panels (a) and (b): Fractions of different types of magnetic field lines; Panels (c) and (d): Fractions of different types of threads; Panels (e) and (f): Scatter plots of the field-line length versus the mean thread length; Panels (g) and (h): Scatter plots of the effective gravity of the dip versus the thread length. The blue solid circles denote the single-dipped cases, and the red solid circles denote the multi-dipped cases. The dashed lines marked in blue, red and green display the fittings to single-dipped cases, multi-dipped cases and all data points, respectively. \label{fig5}}
\end{figure*}

As mentioned in Section \ref{sec:NR1}, although the mean twist is 1.07 turns in the Low-T model and 2.64 turns in the High-T model, both models have non-dipped field lines, single-dipped field lines, and multi-dipped field lines. Figures \ref{fig5}a and \ref{fig5}b show the fractions of the field lines without dips, with a single dip and with multiple dips in the two models. One can see that whereas most field lines (78\%) are non-dipped in the Low-T model, only 26.4\% of the field lines are non-dipped in the High-T model. In particular, 48.4\% of the field lines possess more than one dip in the High-T model. We also calculate the fractions of different types of threads, which are shown in Figs. \ref{fig5}c and \ref{fig5}d. It is found that 82.8\% of the threads are dynamic threads in the Low-T model, and others (17.2\%) are quasi-static threads oscillating near magnetic dips. On the other hand, 60.6\% of the threads are quasi-static threads in the High-T model.

Thread length is another observational feature that can shed light on the magnetic structure of the supporting field lines of prominences \citep{Karpen2003, Zhou2014, Guojh20212}. Rather than investigating the relationship between the thread length and the parameters of the magnetic dip as done by \citet{Zhou2014}, we intend to check the relationship between the thread length and the total length of the magnetic field line for simplicity. For this purpose, we select the quasi-static threads, and the thread length in the multi-dip field lines is the averaged length of multiple threads. In Figs. \ref{fig5}e--\ref{fig5}f, we display the dependence of the thread length ($L_t$) on the field-line length ($L_f$). It shows that despite scattering there is a tendency that the thread length decreases with the field-line length. We divide the data points into single-dipped threads (blue) and multi-dipped threads (red), and then fit the $L_t$--$L_f$ relationships with power-law functions, $L_t=A L_f^B$, for the three groups of data, that is, single-dipped threads (blue),  multi-dipped threads (red) and all threads (green). For the Low-T model in Fig. \ref{fig5}e, the corresponding fitting functions are $L_t=1177.81L_f^{-0.80}$, $L_t=15432.47L_f^{-1.25}$, and $L_t=1430.83L_f^{-0.84}$, respectively, where $L_t$ and $L_f$ are in units of Mm. And the corresponding Spearman correlation coefficients are -0.51, -0.50 and -0.60, respectively. Regarding the High-T model in Fig. \ref{fig5}f, the corresponding fitting functions are $L_t=1417689.60L_f^{-1.92}$, $L_t=2857.02L_f^{-0.87}$ and $L_t=26932.11L_f^{-1.24}$, with the correlation coefficients of -0.83, -0.79 and -0.89, respectively.

It is well known that for a vertical flux tube, the typical length of a plasma structure is the scale height, which is related to gravity. By analogy, the length of a prominence thread along a magnetic dip might be related to effective gravity, which is defined as $g_e = \int_0^{S_d} |g_{\parallel}(s)|{\rm d}s/(S_d g_{\odot})$. As demonstrated by \citet{Guojh20212}, such a single parameter can reflect the magnetic dip configuration comprehensively. Figures \ref{fig5}g--\ref{fig5}h reveal that the thread length decreases with the effective gravity, meaning that longer threads are likely to exist in longer and shallower magnetic dips, which is consistent with \citet{Zhou2014}. The relationship between the thread length and the effective gravity $g_e$ is fitted with a linear function. In the Low-T model, the corresponding fitting functions are $L_{t}=-17.12g_{e}+22.71$, $L_{t}=-31.80g_{e}+25.97$, and $L_{t}=-24.81g_{e}+24.84$ for the single-dipped threads, multi-dipped threads, and all threads, respectively. The corresponding correlation coefficients are -0.64, -0.92 and -0.73, respectively. In the High-T model, the fitting functions are $L_{t}=-45.11g_{e}+31.54$, $L_{t}=-101.55g_{e}+53.64$, and $L_{t}=-73.69g_{e}+40.85$ for the single-dipped threads, multi-dipped threads, and all threads, respectively. The corresponding correlation coefficients are -0.89, -0.77 and -0.86, respectively. We find that the thread length is better correlated with the effective gravity than the field-line length in the Low-T model, but it is opposite for the High-T model. The reason is that in the High-T model, magnetically-connected threads account for a significant proportion. Mutual interactions among the multiple dips along an individual field line reduce the role of the local dip geometry.

\section{Discussions} \label{sec:dis}

\subsection{Morphological differences between prominences with different twists}

Magnetic twist is an important parameter of magnetic structures, describing how many turns a bunch of field lines wind around an axis \citep{LiuR16}. It plays a significant role in predicting solar eruptions and understanding the dynamics of coronal mass ejections \citep[CMEs;][]{Chen2011}. However, the estimate of magnetic twist is non-trivial because it relies on coronal magnetic extrapolations, which are an ill-posed boundary-value problem based on noisy vector magnetic field on the solar surface. Since prominences are usually regarded as the tracers of coronal sheared/twisted magnetic structures, their morphology might reflect the twisting properties of their magnetic structures. In this paper, we explored the morphological differences between two simulated prominences with different twists. The morphological features include (1) how many threads are quasi-static and how many threads are dynamic; (2) the thread length; and (3) whether the prominence is detached from the solar surface.

Observations indicated that some threads are quasi-static, oscillating near the equilibrium positions \citep{Jing2003}, and other threads are dynamic, moving from one polarity to the other as siphon flows \citep{Wang1999, zoup16}. In our simulations, both types of threads exist in the simulated prominences. In the flux rope system with a low twist, the percentage of quasi-static threads is about 17\%, with the remaining 83\% being dynamic threads. On the contrary, in the flux rope system with a high twist, the percentage of quasi-static threads, including independently trapped threads and magnetically connected threads, is about 61\%, with the remaining 39\% are dynamic threads. It is seen that the prominences supported by low-twist flux ropes are mainly composed of short-lived dynamic threads, while the prominences supported by high-twist flux ropes are mainly composed of long-lived quasi-static threads. Therefore, the counterstreamings in a weakly twisted flux rope are composed mainly of alternative unidirectional flows \citep{Zou2017}, while the counterstreamings in a highly twisted flux rope are composed mainly of oscillating threads around magnetic dips \citep{chen14, Zhou2020}. In fact, both types of threads exist in prominences, so the counterstreamings of prominences might be composed of both prominence longitudinal oscillations and unidirectional flows, albeit the proportion is determined by the twist of the supporting flux rope. Besides, in the prominence supported by a high-twist flux rope, there are many short magnetically connected thread pairs, that is, threads pairs connected by double-dipped field lines, which tend to present drastically decaying and decayless oscillations for one thread and the other, respectively \citep{Zhou2017, Zhang2017}.

The thread length is also systematically different in flux ropes with different twists. First, the magnetically connected threads that exist widely in prominences supported by the high-twist flux rope are usually shorter than independently trapped threads. As a result, the vertical piling of these short threads might resemble the observed vertical-like threads in quiescent prominences \citep{Berge2008, Mackay2010, schm10, Schmieder2014}. Second, while shorter dips host shorter threads \citep{Zhou2014}, we found that magnetically connected threads tend to be much shorter. As seen from the synthesized EUV images in Fig. \ref{fig4}d, some longer threads stand out from shorter threads, and these longer threads manifest as prominence barbs. Similar to the dynamic barbs \citep{ouy20}, these barbs do not correspond to prominence feet that extend down to the solar surface.

Observations revealed that some prominences are totally suspended in the corona, whereas others possess two or multiple legs extending down to the solar surface. Our simulations indicate that the prominence in the high-twist flux rope is nearly attached to the solar surface, but the prominence in the low-twist flux rope is somewhat detached to the solar surface. Our results might imply that those detached prominences are probably supported by low-twist flux ropes, and those attached prominences with two endpoints near the solar chromosphere are supported by high-twist flux ropes. The simulation results in our Low-T model are similar to those in \citet{Luna2012}, whose magnetic twist is about 1 turn. In both papers, the threads are almost horizontal, and the prominences are detached from the solar surface. It might imply that there is no sharp dividing line between sheared arcades and weakly-twisted flux ropes.

When viewed along the spine direction, some prominences show horn-like structures, which are intimately related coronal cavities \citep{schm13, Schmit&Gibson2013}. Our simulations well reproduced the horn-like structures. Comparing the end views of the low-twist and high-twist flux ropes in Figs. \ref{fig4}c and \ref{fig4}f, we can find that the horns are much more evident in the Low-T model than in the High-T model. We tentatively suggest that the clearer prominence horns are a signature of less twisted magnetic field lines.

Despite all these differences, the two models reveal some common features in the synthesized prominences. For example, the deviation angle between the threads and the prominence axis ranges from $10^{\circ}$ to $35^{\circ}$, which are consistent with observations \citep{hana17}. On the other hand, our simulations show that such a deviation angle increases from the middle of the prominence spine to the two endpoints, which seems not a ubiquitous feature in observations. It seems that the variation of the deviation angle depends on the magnetic configuration model, and the monotonic variation from the middle of the spine to the endpoints might be an intrinsic property of the TDm flux rope model, which has a low-twist core and a high-twist outer-shell \citep{Guojh20211}. However, it gives us a hint that we might be able to utilize the distribution of the thread deviation angle to diagnose the radial distribution of the magnetic twist in the flux rope. Since the twist profile may affect the threshold of the kink instability \citep{Baty2001}, the distribution of the thread deviation angles can serve as an important proxy of future space weather forecasting.

\subsection{Relationship between prominences and flux ropes}

A statistical study indicated that whereas $\sim$11\% of prominences are supported by sheared magnetic arcades, $\sim$89\% are supported by flux ropes \citep{Ouyang2017}. Hence, in the majority of cases the magnetic structure of prominences is a flux rope, in particular for the quiescent prominences. However, the spatial relationship between a flux rope and a prominence is unclear \citep{Zhouzj2017}, and it was generally assumed that the dense plasmas of a prominence are situated at the magnetic dips, hence are located on the underside of a flux rope \citep{rust94}.

\begin{figure}
  \includegraphics[width=10cm,clip]{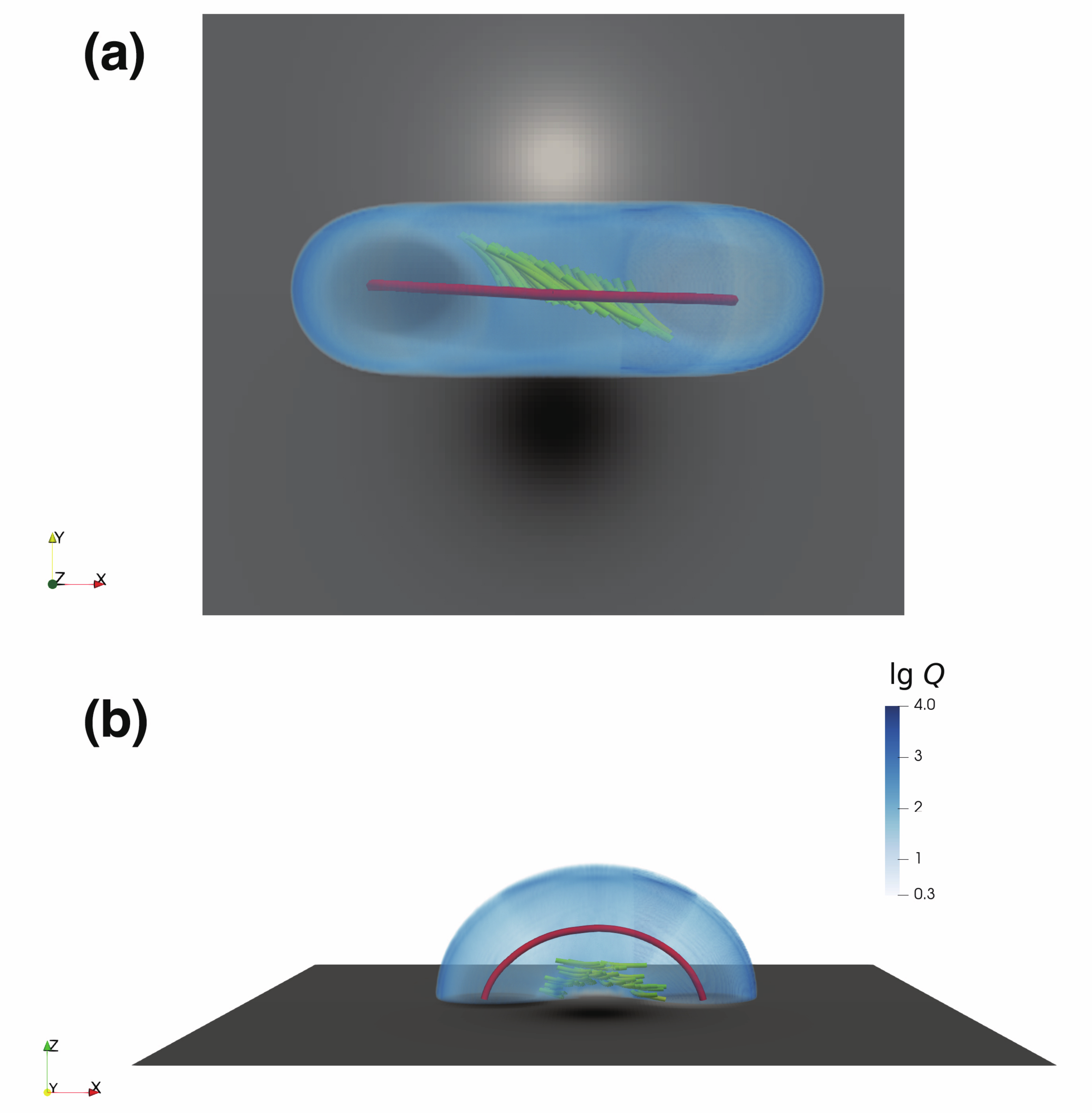}
  \caption{Simulated prominence condensations and the QSL that wraps the flux rope in the Low-T model, where the green isosurface represents the simulated prominence, the blue isosurfaces represent the flux ropes and the red lines indicate the axes of flux ropes. Panels (a) and (b) represent the top and side views, respectively. \label{fig6}}
\end{figure}

In order to check the spatial relationship between prominences and flux ropes, we overlay the simulated prominence with the magnetic flux rope in the Low-twist model together in Fig. \ref{fig6}. It is noted in this figure that the prominence is characterized by the plasma whose temperature is below 20000 K, and the outer boundary of the flux rope is determined by calculating the squashing factor of magnetic connectivity ($Q$ value). Here the outer boundary of the flux rope corresponds to $Q\gg 2$ \citep{Priest1995,Titov2002}, that is, it corresponds to magnetic quasi-separatrix layers (QSLs). It is seen from Fig. \ref{fig6} that not only the prominence deviates from the flux-rope axis, but also the prominence condensations do not fully fill the lower half of the flux rope. It is revealed that in the Low-twist model, the dense plasmas of the prominence occupy only the lower one-fourth in size of the flux rope. Therefore, we have to be careful when taking a prominence as the tracer of a magnetic flux rope for a prominence.

\begin{figure*}
    \includegraphics[width=10cm,clip]{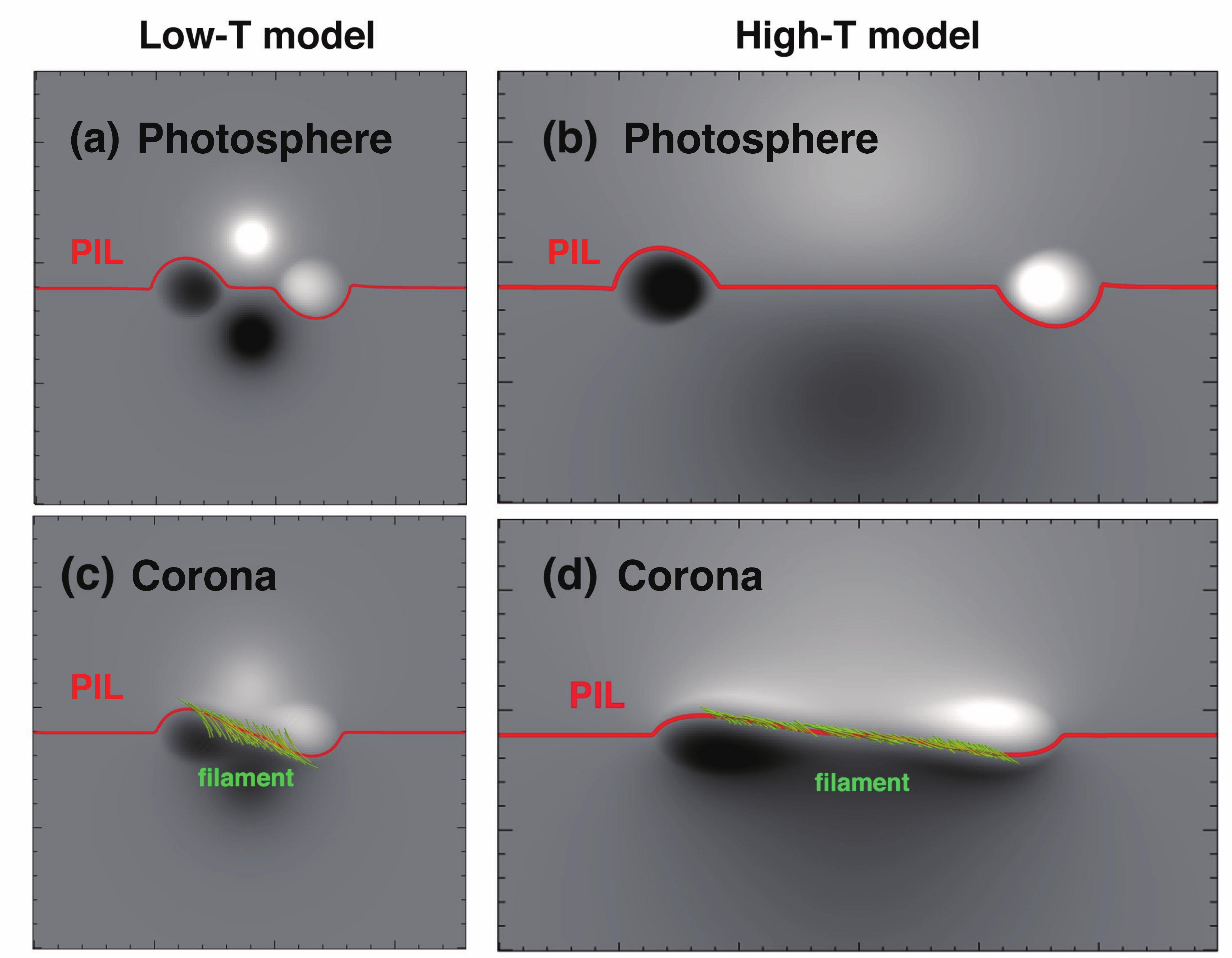}
  \centering
  \caption{Spatial relationship between the PILs and filaments in the Low-T  (panels a and c)and High-T (panels b and d) models, where the red solid lines denote the PILs, and the green isosurfaces represent the simulated filaments, which are overlaid on the magnetograms (gray scale). The first row illustrates the photospheric magnetograms, and the bottom row illustrates the coronal magnetograms at the bottom of the flux rope (10 Mm for the Low-T model and 31 Mm for the High-T model). \label{fig7}}
\end{figure*}

Another caveat is that prominences are often claimed to be located above the magnetic polarity inversion line of the photospheric magnetogram \citep{Mackay2010}. However, according to our simulations, prominence spines are skewed from the photospheric PILs and cospatial with the coronal PILs, as shown in Fig. \ref{fig7}. It should be pointed out that the filament threads in our simulations are located in magnetic dips. However, the troughs of these dips correspond to the local PIL at the bottom of the flux rope in the corona, not the PIL on the photosphere. Depending on the complexity of the magnetic configuration, the coronal PILs and the photospheric PILs might be roughly cospatial or might be skewed significantly.

\section{Summary} \label{sec:sum}

In this paper, we explored the differences of prominence fine structure characteristics in two flux ropes with different twists, one with low twist, and the other with high twist. To summarize, our simulations lead to the following results:

\begin{enumerate}
\item{The types of threads are different in the two models, which might produce different dynamic behaviors. In the low-twist model, the majority of threads are short-lived dynamic threads ($83 \%$), forming high-speed flows inside the filament. However, for the high-twist model, quasi-stationary threads accounts for about $61 \%$, which generally present longitudinal oscillations around the dips.}

\item{The characteristics of the thread length are different in two models, which might produce morphological differences. First, the piling of short magnetically connected threads that widely exist in prominences supported by high-twist flux ropes might resemble the observed vertical-like structures. Second, elongated threads in single-dipped field lines probably stand out from short magnetically connected threads, manifesting as filament barbs without feet.}

\item{The filament spine is not cospatial with its supporting flux rope axis and PIL in the photospheric magnetogram, especially for the low-twist flux rope model. It is found that the prominence not only deviates from the flux rope axis, but also the prominence condensations do not fully fill the lower half of the flux rope. Only the lowest one-forth radial extent of the flux rope is filled with cold plasmas in our simulations. Therefore, one has to be careful when taking a prominence as the tracer of a magnetic flux rope.}
\end{enumerate}

It is noted that our simulations also have some drawbacks. First, these simulations are based on the evaporation-condensation model, whereas other models of prominence formation might influence the appearance of the simulated prominence. For example, for the injection model, cold material injected from the chromosphere is likely to experience expansion, which might form elongated threads \citep{Huang2021}. Second, our simulation results strongly depend on the magnetic configuration of TDm model. In the future, we will consider the pseudo-3D model based on the NLFFF extrapolations and data-driven models from observations. Third, we ignored the effects of the prominences on the magnetic structures. However, the magnetic field might be deformed by the prominence weight when plasma $\delta$ (the ratio of the gravity to the magnetic pressure) is large enough \citep{Zhou2018}. In reality, the heating at the footpoint probably be very complex, for example, the turbulent heating \citep{Zhou2020} or the heating related to the magnetic fields \citep{Yang2018}. However, in this paper, to emphasize the effects of the magnetic configuration, we only considered a simplified and robust pattern (continuous and steady heating independent of the magnetic fields). Nevertheless, the above findings deepened our understanding on the relationships between prominences and their supporting magnetic structures, which provides a scientific basis for studying the magnetic structures of prominences before the eruption. We also expect that high-resolution 3D full MHD simulations in the future can reinforce our conclusions.

\begin{acknowledgements}
This research was supported by National Key Research and Development Program of China (2020YFC2201200), NSFC (12127901, 11961131002, 11773016, and 11533005), and Belgian FWO-NSFC project G0E9619N. The numerical calculations in this paper were performed in the cluster system of the High Performance Computing Center (HPCC) of Nanjing University.
\end{acknowledgements}

\bibliographystyle{aa}
\bibliography{reference}
\end{document}